\documentclass[conference]{IEEEtran}

\usepackage{cite}

\ifCLASSINFOpdf
 \usepackage[pdftex]{graphicx}

\else

\fi

\begin{document}

\title{Is it Possible to Disregard Obsolete Requirements? \\
--- An Initial Experiment on a Potentially New Bias in Software Effort Estimation \\}

\author{\IEEEauthorblockN{Lucas Gren}
\IEEEauthorblockA{Chalmers University of Technology and \\The University of Gothenburg\\
Gothenburg, Sweden\\
Email: lucas.gren@cse.gu.se}
\and
\IEEEauthorblockN{Richard Berntsson Svensson and Michael Unterkalmsteiner}
\IEEEauthorblockA{Software Engineering Research Lab Sweden \\ Blekinge Institute of Technology\\
Karlskrona, Sweden\\
Email: richard.berntsson.svensson@bth.se, \\michael.unterkalmsteiner@bth.se}

}

\maketitle

\begin{abstract}
Effort estimation is a complex area in decision-making, and is influenced by a diversity of factors that could increase the estimation error. The effects on effort estimation accuracy of having obsolete requirements in specifications have not yet been studied. This study aims at filling that gap. A total of 150 students were asked to provide effort estimates for different amounts of requirements, and one group was explicitly told to 
disregard some of the given requirements. The results show that even the extra text instructing participants to exclude requirements in the estimation task, had the subjects give higher estimates. The effect of having obsolete requirements in requirements specifications and backlogs in software effort estimation is not taken into account enough today, and this study provides empirical evidence that it possibly should. We also suggest different psychological explanations to the found effect. 
\end{abstract}


\IEEEpeerreviewmaketitle

\section{Introduction}\label{intro}

One of the most common effort estimation techniques, also in the software industry, is ``expert judgment'' \cite{jorgensen2004review}, which means that an expert estimates what (s)he thinks is the most probable implementation effort (usually in person hours or weeks) for the suggested feature or user story, based on previous experience. These experts' judgments are sometimes inaccurate and studies have shown that unconscious cognitive processes can systematically increase the estimation error in some situations \cite{heuristics}. 

According to a study by Wnuk et al. \cite{wnuk2013obsolete} obsolete 
requirements are very common in practice. In an industry sample of $N=219$, 21.9\% stated that when obsolete requirements are found they: ``move them to a separated section titled obsolete 
requirements or the like,'' 8.9\% ``keep them but assign them a status called 
`obsolete,''' and 54\% ``keep them but assign them a status called `obsolete' 
and write the reason why they became obsolete for future reference.'' This 
clearly points at the fact that extra, and most likely irrelevant, information 
in form of obsolete requirements is visible in requirements specifications and 
backlogs to a large extent in practice today  \cite{wnuk2013obsolete}.

To the best of our knowledge there are not any larger empirical studies that investigate the impact of obsolete requirements on software effort estimation, at least not explicitly or in the public domain. Therefore, the objective of this study is to assess if obsolete requirements, that are explicitly stated to be excluded from the effort estimates, have any impact on the estimates. To investigate this, this study combines person hour expert-judgment estimation with the issue of obsolete requirements in an experimental setting to see if and how such requirements might affect estimation.

We also offer two possible explanations to the found effect; the representativeness heuristic and the decoy effect. 

The representativeness heuristic is when people use a representative example of a previous experience to guide their decision, when, in fact, this experience's similarity to the problem at hand is completely disconnected to its solution \cite{representativeness}. 

The Decoy Effect (or Asymmetric Dominance Effect) is when the presence of a third option in a choice set affects a binary choice. For example, if a person is choosing between buying an apple or a pear (and would have chosen the pear), the mere presence of an orange on sale makes the person to choose the apple instead. The orange is in this case asymmetrically dominated by the other fruit (by quality, price, or preference), but still affects the binary choice (hence the name). It might be desirable if such an effect did not exist since this implicates that there is no best binary choice in many contexts. The Decoy Effect has been proven many times (first by \cite{huber}) and is well-used in marketing today \cite{tversky1,zang}. In addition to effect on choices, \cite{ari1} showed that decision-makers seek subjective dominance to simplify complex decision problems, which explains partly why decoys are so influential.

Through statistical analysis on a large sample ($N=150$) with 
university students, we show that even the extra text instructing participants to 
exclude requirements in the estimation task, had the subjects give 
higher estimates. We therefore also show that obsolete requirements could unfavorably 
trigger students to make large systematic errors that could be avoided in 
software effort estimation.

The remainder of this paper is organized as follows. We describe related 
work to our study in Section~\ref{sec:rw} and the method used for the 
experiment in Section~\ref{sec:method}. We present results in 
Section~\ref{sec:res}, discuss them in Section~\ref{sec:disc} and finish by 
drawing conclusions and giving recommendations for future work in 
Section~\ref{sec:concl}.

\section{Related Work}\label{sec:rw}

There are a number of studies that empirically investigate the impact of irrelevant information on software cost estimates. In \cite{jorgen2}, the results show that pre-planning effort estimates may have an impact on the detailed planning effort estimates, despite subjects being told that the early estimates are not based on historical data. Furthermore, \cite{jorgen3} report that, despite that the subjects were told that customer expectation is not an indicator of the actual effort, irrelevant information about the customer's expectations may affect the cost estimates. Moreover, \cite{jorgensen} investigated the impact of irrelevant and misleading information on software development effort estimates. The results show that irrelevant information may have a small impact on the effort estimates. Finally, in a study by \cite{aranda}, the results show that information that is clearly marked as irrelevant in a requirement specification may have a large impact on software cost estimates. The results in \cite{aranda} could not be explained by the subjects' experience of cost estimations. In general, the above mentioned studies have shown that introducing irrelevant information may lead to an increased estimation error, but with a small sample sizes of around 20 participants in each study. Studies have also shown that domain expertise \cite{ettenson} and attention ability \cite{marz} can play a role on the effect irrelevant information have on a subject's estimations.

Heuristics are mental shortcuts that help humans deal with complex decision-making situations, but sometimes lure us into systematically giving very wrong answers \cite{heuristics}. We will now present two of these cognitive biases, namely the Representativeness Heuristic and the Decoy effect, as they offer two potential explanations to the effect found in our experiment.

\subsection{Defining the Representativeness Heuristic}\label{sec:ade}
A heuristic is a mental shortcut in decision-making that enables us to make faster decisions based on a less thorough analysis, which lessens our cognitive load. Examples of heuristics are stereotyping, rules of thumb and educated guesses. Such shortcuts are much needed and natural part of human life since we cannot deep-dive into every context of all decisions we need to make. Also, all the needed information is often not available to us when a decision needs to be made, which means that the heuristic techniques help us, since they are based on previous and similar experiences. The problem is that we are sometimes tricked in such situations and systematically give very wrong answers \cite{heuristics}.

One of the most common heuristics is the ``representativeness heuristic,'' 
according to which we give a subjective probability to an event based on (1) 
its similarity in characteristics to the parent population and (2) its 
reflection of the salient features of the process by which it is generated. The 
thesis is that A is judged as more probable than B whenever A appears more 
representative than B \cite{representativeness}. In a theoretic analysis of the 
software effort estimation context, \cite{jorgensenrep} also state that the more 
similar A is to B, the more we think that A will also behave like B, i.e.\ we 
think more properties are similar between the two events than the ones actually 
observed as similar. 

The representativeness heuristic is very useful and accurate in many situations but can be troublesome in some contexts \cite{1982juu}:

\begin{quote}
``As the amount of detail in a scenario increases, its probability can only decrease steadily, but its representativeness and hence its apparent likelihood may increase. The reliance on representativeness, we believe, is a primary reason for the unwarranted appeal of detailed scenarios and the illusory sense of insight that such constructions often provide.''
\end{quote}

\subsection{Defining the Decoy Effect}\label{sec:ade}
In decision theory, the axiom of the independence of irrelevant choices states 
that extra irrelevant alternatives will not affect the choice. Or, if x is an 
item in a set A, and A is a subset of B, the probability of choosing x from B 
cannot exceed the probability of choosing x from A. However, \cite{huber} 
showed that adding other alternatives that are used by the decision-maker for 
comparisons affect the binary choice between other items. They call this effect 
the Asymmetric Dominance Effect (ADE), (or later the Decoy Effect) since asymmetrically dominated 
alternatives still affect the choice. This effect can more easily be explained by an example. Ariely \cite{ariely2008pit} tested the decoy effect with MIT students. The experiment was as follows: A person that wants to buy a magazine subscription is presented the following alternatives A and B:

\begin{center}
    \begin{tabular}{ | l | l | l | p{5cm} |}
    \hline
     & A & B \\ \hline
    Price & 59 & 125 \\ \hline
    Type  & Web  & Web and Print  \\ \hline

    \end{tabular}
\end{center}
68 students chose A and 32 students chose B resulting in a total revenue of 8,012. However, if a third option (a decoy) is added:

\begin{center}
    \begin{tabular}{ | l | l | l | p{2cm} |}
    \hline
     & A & B & C \\ \hline
    Price & 59 & 125 & 125 \\ \hline
    Type & Web & Web and Print & Print \\ \hline
    \end{tabular}
\end{center}
16 student chose option A, 84 students chose B, and zero students chose C. No students preferred C but its presence still had the impact that most people now choose B over A resulting in a total revenue of 11,444 \cite{ariely2008pit}.

The research conducted by \cite{ari1} shows that even without a decoy the 
decision-maker seeks subjective dominance in alternatives in order to simplify 
complex decisions. In an effort estimation process, we believe that the 
behavior behind the asymmetric dominance effect could play a role in how effort 
estimations are done, since this is a complex decision process with little 
information at hand, i.e.\ the extra requirements might function as decoys.

We searched related work for studies that empirically investigate the impact of the Decoy Effect and its underlying mechanisms in the context of software effort estimation. To the best of the authors knowledge, there are no studies on the impact of the Decoy Effect in software effort estimation. However, studies where found on the Decoy (or Asymmetric Dominance) Effect in a diversity of fields, e.g. Human--Computer Interaction (e.g. \cite{webdecoy,webdecoy2}), Business (e.g. \cite{zang}), Psychology (e.g. \cite{sla}), and Biology (e.g. \cite{bateson}).

\cite{zang} presents an agent-based model of consumer purchase decision-making that uses multi-agent simulation to exhibit the emergent decoy effect phenomenon. The results from the simulation show that although the nature of the decoy effect is from the consumers' psychological perspective about price\slash quality trade-off, the consumers' sociological interactions still account for a very important part of the decoy effect. \cite{webdecoy} introduced an approach for identifying and minimizing decoy effects. The results from two user experiments clearly show the impact of decoys on decision-making. In \cite{webdecoy2}, an in-depth analysis of of the influences of decoy effect on decision-making was performed. The results show that decoy effects blur the objectivity in decision-making, but the decoys increase the confidence of the decision-maker. In \cite{sla}, the authors investigated whether the decoy effect could be generalized to situations where decision-makers are required to infer attribute values. The results show that decoy effects may have a strong impact and can be generalized to situations that require attribute-level inferences. \cite{bateson} investigated the effects on preferences of adding a third option for birds choosing between artificial flowers. The results show that the decoy option acted to increase the preference of the option that dominated it.

In general, the above mentioned studies have shown that introducing irrelevant information may lead to an increased estimation error. However, studies have shown that domain expertise \cite{ettenson} and attention ability \cite{marz} can play a role in the effect irrelevant information have on a subject. However, we have not been able to find any empirical study that investigates the impact of the Representativeness Heuristic or the Decoy Effect on software effort estimation. Although there are studies about these biases from other research fields, the results should be considered with care since they are conducted in a context that differs from software effort estimation. Therefore, there is a need to conduct empirical studies in contexts similar to those met by software professionals performing software effort estimation. 

\section{Method}\label{sec:method}

\subsection{Design}\label{}
The aim of this study is to assess if obsolete requirements, that are 
explicitly stated to be excluded from the effort estimates, have any impact on 
the size of these estimates.

Since expert judgment is the most commonly used estimation method in industry \cite{jorgensen2004review}, we decided to use this estimation method for this experiment. 

\begin{figure*}
\centerline{
\includegraphics[scale=0.4]{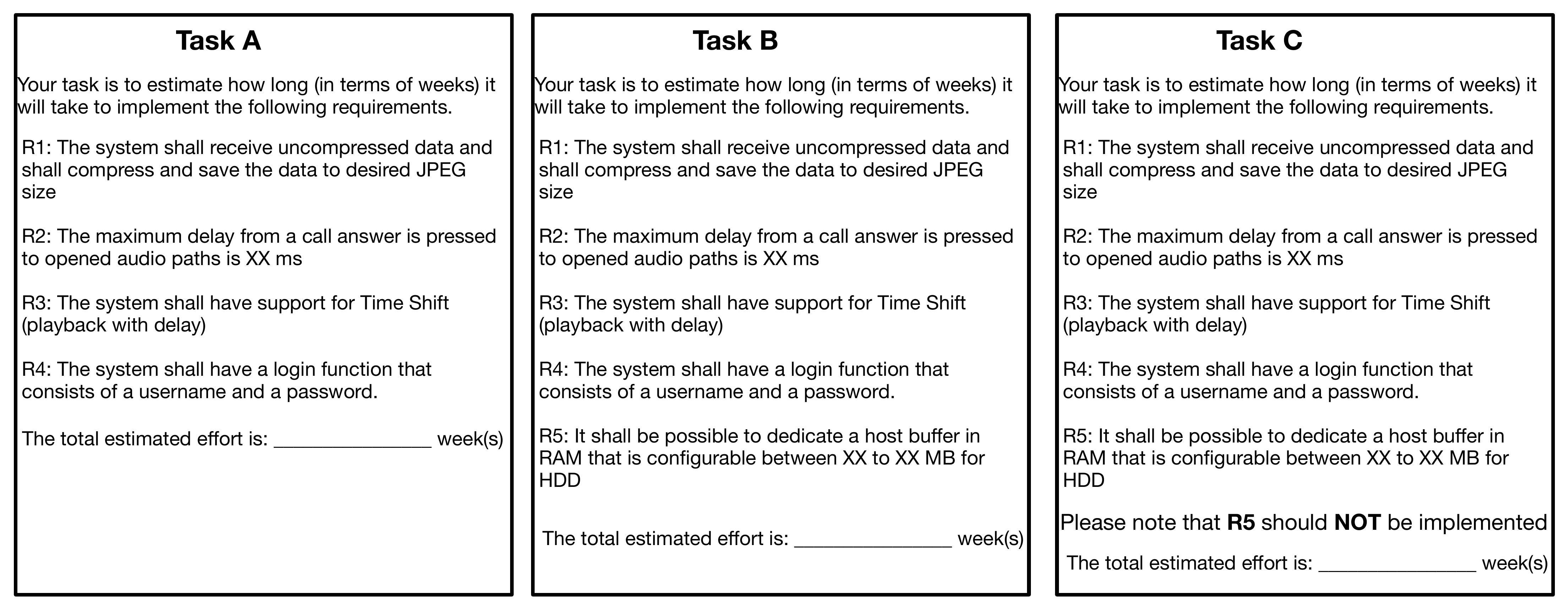}}
\caption{Overview of three requirements specifications (RS1, RS2, and RS3) used in the experiment.}
\label{fig:Tasks}  

\end{figure*}

Three different tasks were designed, Task A, Task B, and Task C. Since the aim of this study is to assess the impact of obsolete requirements in software effort estimations, we needed to have one Requirements Specification (RS) without any obsolete requirements (Task A, which contained four requirements as illustrated in Figure~\ref{fig:Tasks}), and one RS with obsolete requirements (Task C, which contained five requirements as illustrated in Figure~\ref{fig:Tasks}). In addition, since the estimates from Task C may have been similar to, or could have been explained by estimating all of the requirements in Task C, we decided to add a third RS that had the same number of requirements as the RS in Task C, but without any obsolete requirements (Task B). All of the three RSs and their requirements are described in more detail in Section \ref{sec:em}, and shown in Figure~\ref{fig:Tasks}.

\textbf{Research Hypothesis}: We had a preliminary assumption that extra information given in C (and the extra requirements in B) would affect the estimates somehow. Based on this assumption, our hypothesis is that the mean values in groups getting the tasks A, B, and C are different. Or, $H_{1}: \mu _{A} \neq \mu _{B} \neq \mu _{C}$.

Task A was to estimate how long time it would take to implement four requirements from a real industrial project without any obsolete requirements shown. Task B was to estimate the time it would take to implement a total of five requirements from the same industrial project (as in Task A), i.e., one extra requirement as compared to Task A, but the other four were the same. Finally, Task C was to estimate how long time it would take to implement the same five requirements but with the extra text line saying that the last requirement should not be implemented.

\subsection{Subjects}\label{sec:ss}
In this study, the sample included students from the course Software Engineering Process - Economy and Quality (4 ECTS) at Lund University  ($N=150$). The course was a mandatory Bachelor's level course (3rd year students) offered to students at the engineering program Computer Science and Information. The experiment was conducted during a lecture in the course and the results achieved by the subjects had no influence on their final grade. 

Before the experiment started, a pre-questionnaire was used to gather data about the students existing knowledge in terms of the English language, industrial experience in software development, and experience in effort estimations. The result from the pre-questionnaire revealed that the difference in English reading and writing were small, varying from ``very good knowledge'' for the majority of the subjects to ``fluent'' for some of them. When it comes to industrial experience in software development and effort estimation, most of the students reported on no experience at all (116 out of 150, and 147 out of 150). Among the subjects that reported any degree of industrial experience, the length of experience varied between 2 months and one year with an average of 5 months, while the experience for effort estimation varied between 0.5 months and one month.

\subsection{Experimental Material}\label{sec:em}
The material consisted of three short requirement specifications with requirements from a real industrial project from one of our industrial partners in the telecommunication domain that have been implemented, tested, and released. That is, the researchers did not generate requirements for a system for the experiment.

The first requirements specification (RS1 - Task A) consisted of four requirements (R1-R4), e.g.\ \textit{R1: The system shall receive uncompressed data and shall compress and save the data to desired JPEG size}, and \textit{R4: The system shall have a login function that consists of a username and a password}. The task was to estimate how long time it would take to implement these four requirements (see Figure~\ref{fig:Tasks}). The second requirement specification (RS2 - Task B) comprised of five requirements (R1-R5) where the first four (R1-R4) were exactly the same requirements as in RS1. The task was to estimate how long time it would take to implement these five requirements. The third requirement specification (RS3 - Task C) consisted of the same five requirements (R1-R5) as in RS2, but with the additional information that R5 should not be implemented. That is, the task was exactly the same as in RS1. To estimate how long time it would take to implement the requirements in each of there three tasks, the students were asked to estimate the total effort in person-weeks. The reason for using person-weeks was that the requirements from our industrial partner were estimated in person-weeks.

\subsection{Data Collection}\label{sec:dc}
The data was collected as follows. The student subjects were given an introduction and problem description by the moderator (the second author) followed by a pre-questionnaire before starting with the assigned task. After the introduction and the pre-questionnaire, the student subjects were given some time to read through the assignment's instructions. Then, the students completed the estimation work for their assigned task. 

The time estimate in weeks was supposed to be for an individual developer, which was also stated in the introduction given to the subjects. The students were sitting individually when conducting the task and could not interact with each other. In total, the workshop was one hour including introduction, explanations, pre-questionnaire, but the actual time spent on the estimation tasks was 10-20 minutes (even if they where given 25-30 minutes for the task).

Since the requirements are taken from a real system we had to replace the real values with ``XX'' in this paper due to confidentiality reasons.

The task groups A, B, and C were not overlapping, i.e.\ the 150 students were split into three groups for the different estimation tasks (A, B and C).

\subsection{Data Analysis}\label{sec:da}
Since the Kolmogorov-Smirnov tests of normality were significant for the groups Task A and B (Test Statistic for group Task A=0.128, $p$=0.040, Test Statistic for group Task B=0.209, $p$=0.000, and Test Statistic for group Task C=0.108, $p$=0.198), we opted to use the non-parametric Independent-Samples Kruskal-Wallis Test instead of a regular parametric ANOVA. No values were removed from, or altered in, the raw data in this paper. To estimate the effect size we used the $\chi^2$ of the Kruskal-Wallis Test divided by the degrees of freedom \cite{cohen}.

\section{Results}\label{sec:res}
\begin{figure}
\includegraphics[scale=0.54]{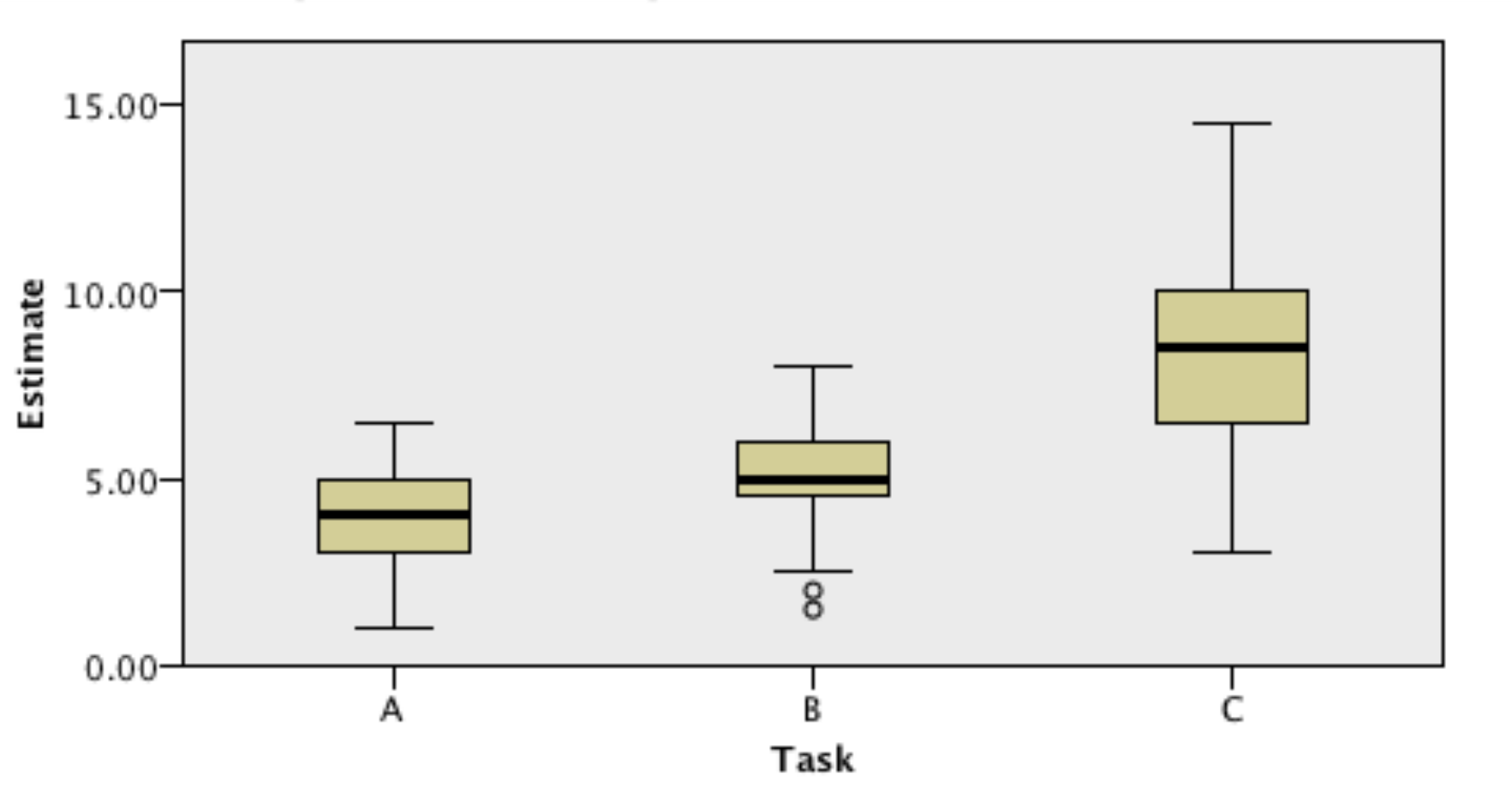}
\caption{Box Plot for Estimates for each Task (A, B, and C).}
\label{fig:boxplot}    
\end{figure}

The conducted Kruskal-Wallis Test can be seen in Table~\ref{fig:kruskal} and the pairwise comparisons between each of the groups are shown in Table~\ref{fig:pairwise}. There were significant differences between all three task groups. To present the result more clearly, Figure~\ref{fig:boxplot} shows box plots for each task. If the subjects were given more requirements the mean estimate increased, as expected. However, when explicitly told to disregard requirements the estimations still significantly increased. It is also important to look at effect size (explained variance by the statistical model) in a statistical test since a significant result could still only explain a very small effect, especially since we have a large sample size. In this case the effect size was $0.485$, which means that 48.5\% of the variance in the estimates can be explained by which task the subjects had. Considering these were different individuals with different experiences (i.e.\ a social science experiment) these are considered very large effects \cite{cohen}. However, such effects have been shown to be smaller in field settings \cite{jorgensen}.

\begin{table}
\caption{Kruskal-Wallis Test (the test statistic is adjusted for ties).}
\label{fig:kruskal}
\begin{center}
    \begin{tabular}{ | l | l | p{5cm} |}
    \hline
    $N$ & 150  \\ \hline
    $\chi^2$ & 72.325 \\ \hline
    $p$ value & 0.000  \\ \hline
    Effect size ($\chi^2$\slash (N-1)) & 0.485  \\ \hline

    \end{tabular}
\end{center}
\end{table}

\begin{table}
\caption{Pairwise group comparisons (2-sided tests and the significance values have been adjusted by the Bonferroni correction for multiple tests).}
\label{fig:pairwise}
\begin{center}
    \begin{tabular}{ | l | l | l | l |p{5cm} |}
    \hline
    \bfseries Group1--Group2 & \bfseries Test Statistic& \bfseries Standard Error & \bfseries p value \\ \hline
    A--B & -23.830 & 8.652& 0.018\\ \hline
    A--C & -72.200 & 8.652 & 0.000\\ \hline
    B--C & -48.370 & 8.652 & 0.000\\ \hline

    \end{tabular}
\end{center}
\end{table}

\section{Discussion}\label{sec:disc}
The results show that an increase of one or two extra requirements has the 
expected (and approximately linear) effect on software effort estimation of 
implementation (Task A compared to Task B). However, to give the extra 
requirement and then explicitly ask the assessor to disregard them (Task C) 
will result in an even higher estimation than if they are left in the 
estimation. This means that this extra ``irrelevant'' information is not only 
taken into account, but leads the assessors to give much higher estimate than if 
stated to be a part of the estimation. This strengthens our confidence that almost all 
types of extra information lead to higher estimations, independently of whether 
the instructions are to include or ignore the information. 

The extra line regarding the exclusion of requirements in the estimate clearly 
confused the subject to a large extent. The statement of exclusion of 
requirements in the specification triggered the subjects to read more 
information into the task then was explicitly there. 

The extra line about the exclusion of the requirement(s) might have 
lured the subjects into comparing and obtaining a more complex system as a 
representative example in their mind, and then they adjust the estimate in 
comparison to that system instead of a smaller system, as the instructions 
imply. If this is true, the representativeness heuristic could be the underlying reason for the estimation bias found. The reason for this is that the extra text then gave more detail to the scenario (the intended system to be build in this case), which meant that the participants were lured into believing that this also gave extra insight that the system was even more complex. As \cite{1982juu} showed, humans show an ``unwarranted appeal of detailed scenarios.'' In this case, the extra text could then have made the requirements specification more representative of a larger system even if the content only was regarding removal of requirements.

Another explanation could be that the mechanisms behind the Decoy Effect is 
what explained the systematic error found. Our experiment was at the individual level of analysis, but the question is if 
the context is possible to disregard. Maybe the representativeness heuristic was not triggered in the participants' minds, but the content on their paper sheets actually had the students in Task group C to, not just estimate the requirements they were 
supposed to, but instead, take the whole requirements specification as well as the setting into account in their estimations.
While the representativeness heuristic might be the more appropriate 
explanation than the decoy effect, further studies are needed to attribute 
their individual impact on effort estimations to find out which one, or possibly both, that accounted for the found effect. \cite{webdecoy2} 
showed that that decoy effects blur the objectivity in decision-making but 
increase the confidence of the decision-maker, which is definitely an aspect 
that should be researched in the software effort estimation context. \cite{sla} 
also showed that decoy effects are present even if not numerical values and the 
whole choice set is presented to the subject. They state that this shows 
evidence that decoy effects are present in many decision-making situations and 
are very generalizable. All in all, these presented findings suggest decoys 
have a large impact on situations very similar to that of a work group or 
individual conducting time estimation for projects.

With a large sample we have reason for generalizing our findings to the larger 
population of students working with requirements as text units. However, we 
would still be very careful when drawing conclusions to practitioners since we 
have a clear ``toy problem'' threat to our study. 
The Decoy Effect (or more specifically the tendency to seek for subjective 
dominance in decision-making) could also play a role when researching\slash 
writing 
requirements, i.e.\ extra information and requirements not needed for the 
estimation will affect the decision-makers assessment of the complexity and 
therefore also the estimates. This might also be a factor to consider in agile 
software development since the product (or sprint) backlog often contains 
irrelevant requirements\slash user stores that have been decided not be 
implemented when estimating the efforts for these different stories. The 
product backlog can include many requirements, but the sprint backlog has a 
smaller amount of stories \cite{schwaber}. This study has shown that the mere 
fact that they are visible to the people doing the estimates could have a vast 
impact on the estimates' accuracy.

\subsection{Limitations}\label{sec:limits}
We would like to highlight that we do not know if any of the cognitive biases presented really explain the effect we found, however, we believe that some of the psychological reasoning might overlap, i.e., we need to run more experiments and get more qualitative data to understand why the students gave higher estimates when exposed to obsolete requirements. 

The largest threat to our study is the artificial experimental setting. None of 
the research subjects had any personal commitment to implementing the 
requirements that where given to them. We tried to mitigate this by giving all 
subjects a thorough introduction to the experiment and ask them to give as 
accurate estimates as possible. The artificial setting might explain the very 
high effects found in our data, but even if the effects are much smaller in 
practice, we argue that these cognitive biases still exist and should be dealt 
with accordingly. Another limitation is that non-intended interpretations of 
the estimation instructions may explain the observed effects. However, for 
whatever reason behind these higher estimates, they are still clearly visible 
in our sample ($N=150$). Also, the students lack expertise in requirements estimation, which means that we might not investigate the use of expert judgment in our experiment. However, we were interested in the relative difference between the estimations given in the three different Task groups A, B, and C and therefore the absolute estimates were not the main focus. We therefore assumed that students all have the same expertise, i.e., their knowledge on the domain and estimation technique is homogeneous, and therefore, differences in effort estimates can only be attributed to the task. We did not look at the accuracy or correctness of the estimates in this study.

\section{Conclusions and Future Work}\label{sec:concl}
Through an experiment including university students ($N=150$), we have found 
that adding obsolete extra requirements could have a large impact on the effort 
estimates. Being instructed to disregard requirements should intuitively 
decrease the estimated implementation time, but this study shows that adding 
such a statement has the opposite effect. These findings are important 
contributions to both researchers and practitioners since obsolete requirements 
might have an impact on the estimation error in software effort estimation. 

A natural next step would be to increase the sample in future studies, collect 
qualitative data on the participants reasoning, and also include industry 
practitioners in order to validate if the effect is still visible in such 
settings. There are undeniably cognitive aspects that influence software effort 
estimation that are currently not taken into enough consideration.

\bibliographystyle{IEEEtran}

\bibliography{refs}

\begin{thebibliography}{10}
\providecommand{\url}[1]{#1}
\csname url@samestyle\endcsname
\providecommand{\newblock}{\relax}
\providecommand{\bibinfo}[2]{#2}
\providecommand{\BIBentrySTDinterwordspacing}{\spaceskip=0pt\relax}
\providecommand{\BIBentryALTinterwordstretchfactor}{4}
\providecommand{\BIBentryALTinterwordspacing}{\spaceskip=\fontdimen2\font plus
\BIBentryALTinterwordstretchfactor\fontdimen3\font minus
  \fontdimen4\font\relax}
\providecommand{\BIBforeignlanguage}[2]{{%
\expandafter\ifx\csname l@#1\endcsname\relax
\typeout{** WARNING: IEEEtran.bst: No hyphenation pattern has been}%
\typeout{** loaded for the language `#1'. Using the pattern for}%
\typeout{** the default language instead.}%
\else
\language=\csname l@#1\endcsname
\fi
#2}}
\providecommand{\BIBdecl}{\relax}
\BIBdecl

\bibitem{jorgensen2004review}
M.~J{\o}rgensen, ``A review of studies on expert estimation of software
  development effort,'' \emph{Journal of Systems and Software}, vol.~70, no.~1,
  pp. 37--60, 2004.

\bibitem{heuristics}
A.~Tversky and D.~Kahneman, ``Judgment under uncertainty: Heuristics and
  biases,'' \emph{Science}, vol. 185, no. 4157, pp. 1124--1131, 1974.

\bibitem{wnuk2013obsolete}
K.~Wnuk, T.~Gorschek, and S.~Zahda, ``Obsolete software requirements,''
  \emph{Information and Software Technology}, vol.~55, no.~6, pp. 921--940,
  2013.

\bibitem{representativeness}
D.~Kahneman and A.~Tversky, ``Subjective probability: A judgment of
  representativeness,'' in \emph{The Concept of Probability in Psychological
  Experiments}.\hskip 1em plus 0.5em minus 0.4em\relax Springer, 1974, pp.
  25--48.

\bibitem{huber}
J.~Huber, J.~W. Payne, and C.~Puto, ``Adding asymmetrically dominated
  alternatives: Violations of regularity and the similarity hypothesis,''
  \emph{Journal of consumer research}, pp. 90--98, 1982.

\bibitem{tversky1}
A.~Tversky and I.~Simonson, ``Context-dependent preferences,'' \emph{Management
  science}, vol.~39, no.~10, pp. 1179--1189, 1993.

\bibitem{zang}
T.~Zhang and D.~Zhang, ``Agent-based simulation of consumer purchase
  decision-making and the decoy effect,'' \emph{Journal of business research},
  vol.~60, no.~8, pp. 912--922, 2007.

\bibitem{ari1}
D.~Ariely and T.~S. Wallsten, ``Seeking subjective dominance in
  multidimensional space: An explanation of the asymmetric dominance effect,''
  \emph{Organizational Behavior and Human Decision Processes}, vol.~63, no.~3,
  pp. 223--232, 1995.

\bibitem{jorgen2}
M.~J{\o}rgensen and D.~I. Sj{\o}berg, ``Impact of effort estimates on software
  project work,'' \emph{Information and Software Technology}, vol.~43, no.~15,
  pp. 939--948, 2001.

\bibitem{jorgen3}
------, ``The impact of customer expectation on software development effort
  estimates,'' \emph{International Journal of Project Management}, vol.~22,
  no.~4, pp. 317--325, 2004.

\bibitem{jorgensen}
M.~J{\o}rgensen and S.~Grimstad, ``The impact of irrelevant and misleading
  information on software development effort estimates: A randomized controlled
  field experiment,'' \emph{Software Engineering, IEEE Transactions on},
  vol.~37, no.~5, pp. 695--707, Sept 2011.

\bibitem{aranda}
J.~Aranda and S.~Easterbrook, ``Anchoring and adjustment in software
  estimation,'' in \emph{Proceedings of the 10th European Software Engineering
  Conference Held Jointly with 13th ACM SIGSOFT International Symposium on
  Foundations of Software Engineering}, ser. ESEC/FSE—-13.\hskip 1em plus
  0.5em minus 0.4em\relax New York, NY, USA: ACM, 2005, pp. 346--355.

\bibitem{ettenson}
R.~Ettenson, ``\BIBforeignlanguage{English}{Expert judgment: is more
  information better?}'' \emph{\BIBforeignlanguage{English}{Psychological
  reports}}, vol.~60, no.~1, p. 227–238, 1987.

\bibitem{marz}
G.~M. Marzocchi, D.~Lucangeli, T.~De~Meo, F.~Fini, and C.~Cornoldi, ``The
  disturbing effect of irrelevant information on arithmetic problem solving in
  inattentive children,'' \emph{Developmental Neuropsychology}, vol.~21, no.~1,
  p. 73–92, 2002.

\bibitem{jorgensenrep}
M.~J{\o}rgensen and D.~Sj{\o}berg, ``The importance of not learning from
  experience,'' in \emph{Proc. European Software Process Improvement Conf},
  2000, p. 2–10.

\bibitem{1982juu}
D.~Kahneman, P.~Slovic, and A.~Tversky, \emph{Judgement under uncertainty:
  Heuristics and biases}.\hskip 1em plus 0.5em minus 0.4em\relax Cambridge:
  Cambridge U.P., 1982.

\bibitem{ariely2008pit}
D.~Ariely, \emph{Predictably irrational: {T}he hidden forces that shape our
  decisions}.\hskip 1em plus 0.5em minus 0.4em\relax London: HarperCollins,
  2008.

\bibitem{webdecoy}
E.~Teppan and A.~Felfernig, ``Minimization of product utility estimation errors
  in recommender result set evaluations,'' in \emph{Web Intelligence and
  Intelligent Agent Technologies, 2009. WI-IAT '09. IEEE/WIC/ACM International
  Joint Conferences on}, vol.~1, Sept 2009, pp. 20--27.

\bibitem{webdecoy2}
E.~Teppan, G.~Friedrich, and A.~Felfernig, ``Impacts of decoy effects on the
  decision making ability,'' in \emph{Commerce and Enterprise Computing (CEC),
  2010 IEEE 12th Conference on}, Nov 2010, pp. 112--119.

\bibitem{sla}
J.~E. Slaughter, E.~F. Sinar, and S.~Highhouse, ``Decoy effects and
  attribute-level inferences.'' \emph{Journal of applied psychology}, vol.~84,
  no.~5, p. 823, 1999.

\bibitem{bateson}
M.~Bateson, S.~D. Healy, and T.~A. Hurly, ``Context--dependent foraging
  decisions in rufous hummingbirds,'' \emph{Proceedings of the Royal Society of
  London. Series B: Biological Sciences}, vol. 270, no. 1521, pp. 1271--1276,
  2003.

\bibitem{cohen}
J.~Cohen, ``Quantitative methods in psychology - a power primer,''
  \emph{Psychological Bulletin}, vol. 112, no.~1, pp. 155--159, 1992.

\bibitem{schwaber}
K.~Schwaber, \emph{Agile project management with Scrum}.\hskip 1em plus 0.5em
  minus 0.4em\relax Redmond, Wash.: Microsoft Press, 2004.

\end{thebibliography}

\end{document}